\documentclass[twocolumn,trackchanges]{aastex7}

\usepackage{natbib} 
\usepackage{graphicx}	
\usepackage{amsmath}	
\usepackage{hyperref}
\usepackage{cprotect}
\usepackage{multirow}
\usepackage{placeins} 
\usepackage{orcidlink}

\usepackage{xspace}

\usepackage[version=4]{mhchem}
\usepackage{wrapfig}

\shortauthors{Werlen et al. }

\shorttitle{Atmospheric C/O Ratios of Sub-Neptunes with Magma Oceans}

\begin{document}

\title{Atmospheric C/O Ratios of Sub-Neptunes with Magma Oceans: Homemade rather than Inherited}

\author[orcid=0009-0005-1133-7586, sname='Werlen']{Aaron Werlen}
\affiliation{Institute for Particle Physics and Astrophysics, ETH Zurich, CH-8093 Zurich, Switzerland}
\email[show]{awerlen@ethz.ch}  

\author[orcid=0000-0001-6110-4610, sname='Dorn']{Caroline Dorn}
\affiliation{Institute for Particle Physics and Astrophysics, ETH Zurich, CH-8093 Zurich, Switzerland}
\email{dornc@ethz.ch}

\author[orcid=0000-0002-0298-8089, sname='Schlichting']{Hilke E. Schlichting}
\affiliation{Department of Earth, Planetary, and Space Sciences, University of California, Los Angeles, CA 90095, USA}
\email{hilke@ucla.edu}

\author[orcid=0000-0002-0632-4407, sname='Grimm']{Simon L. Grimm}
\affiliation{Institute for Particle Physics and Astrophysics, ETH Zurich, CH-8093 Zurich, Switzerland}
\affiliation{Department of Astrophysics, University of Zurich, CH-8057 Zurich, Switzerland}
\email{sigrimm@ethz.ch}

\author[orcid=0000-0002-1299-0801, sname='Young']{Edward D. Young}
\affiliation{Department of Earth, Planetary, and Space Sciences, University of California, Los Angeles, CA 90095, USA}
\email{eyoung@epss.ucla.edu}

\begin{abstract}
Recently, the James Webb Space Telescope has enabled detailed spectroscopic characterization of sub-Neptune atmospheres. With detections of carbon- and oxygen-bearing species such as CO, CO$_2$, CH$_4$, and H$_2$O, a central question is whether the atmospheric C/O ratio, commonly used to trace formation location in giant planets, can serve a similar diagnostic role for sub-Neptunes. We use the global chemical equilibrium framework of \citet{schlichting_chemical_2022} to quantify how magma ocean-atmosphere interactions affect the atmospheric C/O ratio. We find that the resulting C/O ratios range from several orders of magnitude below solar to a few times solar. The atmospheric C/O ratio in sub-Neptunes is therefore not inherited from the protoplanetary disk, but instead emerges from chemical equilibrium between the atmosphere and the underlying magma ocean. Planetary mass, atmospheric mass fraction, and thermal state all strongly influence the atmospheric C/O ratio. In addition, carbon partitioning into the metal phase typically reduces the atmospheric C/O ratio substantially, particularly for atmospheric mass fractions less than a few percent. Finally, we couple the deep equilibrium compositions to 1D atmospheric models that self-consistently solve for the pressure-temperature structure and chemical composition, including photochemistry. We find that the C/O ratio varies with altitude under low vertical mixing conditions (K$_\text{zz}=10^4$~cm$^2$~s$^{-1}$), but remains constant under strong mixing (K$_\text{zz}=10^7$~cm$^2$~s$^{-1}$). Our results imply that observed C/O ratios of sub-Neptunes can be used to probe their interiors. Specifically, C/O ratios much lower than host star values would imply an underlying magma ocean with iron metal having sequestered significant amounts of carbon.
\end{abstract}



\section{Introduction}\label{sec:intro}

Efforts to connect atmospheric compositions with planet formation began with retrieval studies of hot Jupiters. \citet{madhusudhan_high_2011} conducted one of the first such studies, inferring a super-stellar atmospheric C/O ratio ($>1$) for WASP-12b based on Spitzer eclipse photometry. However, this result was contested by subsequent analyses, which did not confirm the high C/O ratio \citep{crossfield_re-evaluating_2012, swain_probing_2013, line_systematic_2014, kreidberg_detection_2015}.

In parallel, theoretical studies explored how disk chemistry might influence planetary composition. \citet{oberg_effects_2011} proposed that the C/O ratio of a planet's atmosphere could be inherited from its formation location relative to the protoplanetary disk’s volatile ice lines. In this model, C/O~$\sim$~1 is possible only for planets enriched mainly through gas accretion and formed between the CO and CO$_2$ ice lines. \citet{helling_disk_2014} reached a similar conclusion by modeling the chemical evolution of gas and ices in protoplanetary disks, although they did not simulate planet formation directly.

Later studies incorporated more physically consistent models of planet formation and chemistry. \citet{thiabaud_stellar_2014, thiabaud_gas_2015} and \citet{marboeuf_planetesimals_2014, marboeuf_stellar_2014} modeled planet formation via core accretion, tracking both refractory and volatile element condensation to predict final planetary compositions. \citet{madhusudhan_toward_2014} explored whether observed atmospheric abundances could constrain formation pathways for hot-Jupiters. They found that giant planets forming in the outer disk could have sub-stellar C/O ratios if their enrichment is dominated by gas accretion. If such planets are found at small orbital distances, this would require inward migration after disk dispersal. \citet{mordasini_imprint_2016} extended these approaches by modeling the chemical enrichment of giant planets through planetesimal accretion, finding that atmospheric compositions are largely determined by the properties of accreted solids—consistent with observations of both Solar System and extrasolar planets.

Within the Solar System, formation–composition links have long been used to interpret the atmospheric abundances of giant planets such as Jupiter and Saturn \citep[e.g.,][]{mousis_determination_2009}. More recently, \citet{mousis_primordial_2025} interpreted JWST observations of Eris and Makemake using primordial compositional models.

Until recently, such compositional studies have focused primarily on giant or dwarf planets. However, JWST now enables detailed atmospheric characterization of sub-Neptunes \citep[e.g.,][]{madhusudhan_carbon-bearing_2023, holmberg_possible_2024, benneke_jwst_2024, schmidt_comprehensive_2025, felix_evidence_2025}. Notably, \citet{benneke_jwst_2024} and subsequently \citet{felix_evidence_2025} reported the first measurement of the atmospheric C/O ratio in the 2.2~R$_\oplus$ sub-Neptune TOI-270d, based on transmission spectroscopy.

Given this new observational access to sub-Neptune atmospheres, a key question that arises is whether their compositions are primarily determined by the accretion of ices in the protoplanetary disk—thereby serving as a tracer of formation history—or whether subsequent processes such as magma ocean–atmosphere interactions dominate the atmospheric composition. 

The interior structure and thermal evolution of sub-Neptunes differ fundamentally from those of hot Jupiters and dwarf planets. Planets with a few Earth masses embedded in a gas disk can accrete several weight percent of hydrogen–helium gas \citep[e.g.,][]{lee_cool_2015, ginzburg_super-earth_2016}. During formation, the gravitational binding energy released during a planet's assembly is converted into heat. The resulting temperatures, likely limited by the upper threshold for stable H/He envelopes, can exceed $10^4$~K in super-Earths and sub-Neptunes \citep{ginzburg_super-earth_2016}.

The massive H/He envelope acts as an insulating layer, significantly reducing surface cooling and delaying the solidification of the magma ocean to gigayear timescales \citep[e.g.,][]{lopez_understanding_2014, ginzburg_super-earth_2016}. Under such conditions, the planetary surface remains molten for extended periods, allowing continuous chemical exchange between the atmosphere and the interior. Several studies \citep[e.g.,][]{ginzburg_super-earth_2016, chachan_role_2018, kite_superabundance_2019, schlichting_chemical_2022, kite_water_2021, tian_atmospheric_2024, seo_role_2024} have shown that understanding these atmosphere–interior interactions is essential for interpreting the evolution and atmospheric composition of sub-Neptunes.

In this study, we investigate how chemical equilibrium between sub-Neptune atmospheres and molten interiors may affect the atmospheric C/O ratio, across a range of planetary masses, atmospheric mass fractions, atmosphere–magma ocean interface (AMOI) temperatures, and silicate–metal equilibrium (SME) temperatures. To this end, we employ the global chemical equilibrium framework introduced by \citet{schlichting_chemical_2022}. We extend this framework to allow carbon to partition from the silicate into the metal phase and analyze how this process modifies the resulting atmospheric C/O ratio.

This paper is structured as follows. In Section~\ref{sec:Method}, we describe the global chemical equilibrium framework, outline the extensions made to include carbon partitioning between the silicate and metal phases, and present the grid of planetary masses, atmospheric mass fractions, AMOI temperatures, and SME temperatures used in the simulations. We also describe the coupling to a 1D atmospheric model. In Section~\ref{sec:Results}, we present the resulting trends in atmospheric C/O ratios, including the effects of carbon partitioning between phases, and examine their dependence on planetary parameters. In Section~\ref{sec:Discussion}, we compare our findings to previous studies of magma ocean–atmosphere interaction and discuss implications for interpreting C/O ratios measured by JWST. We summarize our key results and conclusions in Section~\ref{sec:Conclusions}.

\section{Methods}\label{sec:Method}
\subsection{Chemical Thermodynamics}\label{sec:chemical_network}

We make use of a variant of the global chemical equilibrium code of \citet{schlichting_chemical_2022}. The code solves a set of linearly independent reactions that span the relevant reaction space for magma ocean–atmosphere interactions. Our model includes 19 chemical reactions involving 26 phase components. Reactions occur within each of the three phases—metal, silicate, and gas—as well as between them. The core, in the astrophysical sense, is composed of two phases: metal and silicate.

We use this definition to underscore that the metal and silicate phases do not necessarily need to be separated by a well-defined core–mantle boundary. Experimental studies \citep{hirao_compression_2004,terasaki_hydrogen_2009, tagawa_experimental_2021} have found evidence for limitless hydrogen solubility in metallic phases under high-pressure conditions. Ab initio calculations by \citet{li_earths_2020} support these findings theoretically. The chemical equilibrium code used in this study has also yielded large quantities of hydrogen, oxygen, and silicon in the metallic phase \citep{schlichting_chemical_2022, young_earth_2023, young_phase_2024}. Together, these results suggest that a metal phase can harbor large amounts of hydrogen, among other light elements like O and Si.  The resulting densities of these metals are comparable to that of the silicate, potentially precluding separation by gravitational settling \citep{young_phase_2024}.

For this study, we adopt all 18 reactions from \citet{schlichting_chemical_2022} and expand the network by introducing carbon partitioning from the metal to the silicate phase through the following reaction:

\begin{equation}
    \label{Carbonreaction} \ce{C_{metal} + O_{metal} \rightleftharpoons CO_{silicate}}
\end{equation}

\noindent Reaction~\ref{Carbonreaction}, combined with the reaction set in \cite{schlichting_chemical_2022}, defines an example basis set. All linear combinations of this basis set are, by definition, included in the equilibrium reaction network. For example, although the dissolution of CH$_4$ is not explicitly listed, it is captured through a combination of reactions involving CH$_4$ oxidation, CO dissolution, and subsequent carbon partitioning into the metal phase. This means that any pathway involving CH$_4$ is accounted for in the bases set of reactions and is therefore inherently included in the equilibrium calculation.

Chemical equilibrium for the 19 reactions is obtained by solving the following equilibrium condition for all molar mass fractions $x_i$:

\begin{equation}\label{eq:chemical_equilibrium}
    \sum_i \nu_i \ln x_i + \left[\frac{\Delta \hat{G}^\circ_{\text{rxn}}}{RT} + \sum_g \nu_g \ln(P/P^\circ)\right] = 0,
\end{equation}

\noindent where $x_i$ is the mole fraction of species $i$ in its host phase, $\nu_i$ are the stoichiometric coefficients for species $i$, $\Delta\hat{G}^\circ_\text{rxn}$ is the standard-state Gibbs free energy of reaction, $R$ is the ideal gas constant, $T$ is the temperature, $P$ is pressure at the AMOI, and $P^\circ$ is the standard-state pressure (1 bar in this case). The index $i$ spans all species, while $g$ refers to gas-phase species only, for which we include a pressure dependence.

In addition to the equilibrium conditions, we add three equations enforcing mass balance constraints by requiring that the mole fractions in each phase sum to unity:

\begin{equation}
    1 - \sum_i x_{i,k} = 0,
\end{equation}

\noindent where $x_{i,k}$ is the mole fraction of component $i$ in phase $k$.

We impose elemental conservation by introducing seven additional sum constraints—one for each conserved element:

\begin{equation}
    n_s - \sum_i \sum_k n_{s,i,k} x_{i,k} N_k = 0,
\end{equation}

\noindent where $n_s$ is the total number of moles of element $s$, $x_{i,k}$ is the mole fraction of component $i$ in phase $k$, $n_{s,i,k}$ is the number of atoms of element $s$ in component $i$ of phase $k$, and $N_k$ is the total number of moles of phase $k$. In our model, both the mole fractions $x_{i,k}$ and the total moles for phases $k = $ metal, silicate, or gas, $N_k$,  are treated as variables.
The pressure at the AMOI is implied by the derived mean molecular weight of the atmosphere.   

We solve for the global chemical equilibrium following the numerical scheme of \citet{schlichting_chemical_2022}, but with key adaptations that reduce the computational time from approximately 30 minutes to a few tens of seconds on a single CPU. The details of these improvements are presented in Grimm et al. (in prep).

\subsection{Thermodynamic Data}

To solve the chemical equilibrium network, we require the standard-state molar Gibbs free energies of reaction, $\Delta \hat{G}^\circ_\text{rxn}$. These values are calculated following the approach described in \citet{schlichting_chemical_2022}. Where available, we use internally consistent Gibbs free energies of formation for silicate and metal species. Ideal mixing is assumed for all phases except for Si and O in metal \citep{young_earth_2023}. When direct thermodynamic data are not available for a given species, we derive its Gibbs free energy of formation by combining the free energies of known species with known free energies of reaction. 

Following \citet{schlichting_chemical_2022}, we do not apply pressure corrections to the Gibbs free energies of silicate species. This choice is motivated by the expectation that pressure effects for intramelt reactions are minor and may largely cancel out \citep[see][]{stixrude_structure_2005,de_koker_self-consistent_2009,schlichting_chemical_2022}. Nonetheless, additional thermodynamic constraints are needed to more robustly assess these effects. Similarly, no pressure corrections are applied to metal species. For the detailed derivation of the thermodynamic data and reaction framework, we refer the reader to the appendix in \citet{schlichting_chemical_2022}.

\subsection{Siderophile Behavior of Carbon}

We rely on experimentally determined partitioning of carbon between metal and silicate phases in order to evaluate the free energy of reaction \ref{Carbonreaction}. We assume that the concentration of carbon in the silicate, expressed as x$_{\text{CO,silicate}}$, is equivalent to x$_{\text{C,silicate}}$ in the experiments. Experimental measurements of carbon partitioning vary significantly, particularly in their pressure dependence, which is of key importance in the present context. \citet{grewal_delivery_2019} report an increase in the siderophile (iron-loving) behavior of carbon with increasing pressure, whereas \citet{fischer_carbon_2020} find the opposite trend.

For this study, we adopt the carbon partitioning behavior from \citet{blanchard_metalsilicate_2022}, who used $^{13}$C to avoid potential contamination from the diamond anvil cell. In addition, \citet{blanchard_metalsilicate_2022} excluded large quenched metal blebs from their silicate analyses, which may not reliably reflect high-pressure equilibrium conditions. In contrast, \citet{fischer_carbon_2020} included such blebs, interpreting them as quench artefacts and assuming they would have been dissolved at high pressure and temperature. \citet{blanchard_metalsilicate_2022} also noted that they were not able to reproduce the high partitioning fits reported by \citet{grewal_delivery_2019}. For these reasons, we adopt the partitioning data from \citet{blanchard_metalsilicate_2022}. In our model, this yields carbon partitioning behavior intermediate between that of \citet{fischer_carbon_2020} and that of \citet{grewal_delivery_2019}.

In order to assess the role of carbon partitioning to the metal phase, we run the entire planetary grid both with the full chemical network described in Section~\ref{sec:chemical_network} and with the original reaction network from \citet{schlichting_chemical_2022}, which excludes both reaction~\ref{Carbonreaction} and carbon in the metal phase. This comparison allows us to isolate and quantify the effect of carbon in the metal phase on the atmospheric C/O ratio.

\subsection{Linking Deep Atmosphere with Observable Upper Atmosphere}\label{sec:observable_atmosphere}

To assess how the chemical composition of the deep atmosphere, as set by equilibrium with the magma ocean, is transmitted to the observable upper layers of the atmosphere, we employ a suite of open source atmospheric models: \texttt{FastChem} \citep{stock_fastchem_2018}, \texttt{HELIOS} \citep{malik_helios_2017, malik_self-luminous_2019}, \texttt{VULCAN} \citep{tsai_vulcan_2017}, and \texttt{HELIOS-K} \citep{grimm_helios-k_2021}. These models together enable the simulation of chemical equilibrium, photochemistry, and radiative transfer with species-dependent opacities over a broad pressure range.

\texttt{FastChem} computes gas-phase chemical equilibrium under local thermodynamic conditions. \texttt{HELIOS} is a one-dimensional radiative-convective model used to compute pressure-temperature (P–T) profiles. \texttt{HELIOS-K} calculates wavelength-dependent opacities for given gas mixtures. \texttt{VULCAN} is a solver for kinetic reaction networks that incorporates thermochemistry, photochemistry, vertical mixing, and condensation. Together, these models allow us to simulate atmospheric structures from the magma ocean interface to atmospheric pressures of $10^{-8}$~bar.

Using these codes, our workflow is as follows. We first calculate the atmospheric composition above the magma obtained using our global equilibrium model. We then use \texttt{FastChem} to define the molar equilibrium mixing ratios of gas species for this composition for a P-T grid representing possible atmosphere conditions above the magma ocean. This output is passed to \texttt{HELIOS-K} to compute gas opacities, which in turn serve as inputs for \texttt{HELIOS} to derive the atmosphere P–T profile. The P-T profile, along with the gas composition above the magma ocean, comprising the lower boundary condition, is fed into \texttt{VULCAN} to compute the steady-state chemical structure. We use the output from \texttt{VULCAN} to  recompute the opacities with \texttt{HELIOS-K} and update the P–T structure with \texttt{HELIOS}. This iterative cycle between \texttt{HELIOS}/\texttt{HELIOS-K} and \texttt{VULCAN} is repeated until convergence is achieved in both the mixing ratios and the P-T profile. The pressure grid in \texttt{HELIOS} extends from the AMOI pressure up to $10^{-5}$~bar. In \texttt{VULCAN}, we assume perfect mixing in the deep convective layer and begin calculating chemical kinetics from
$10^{3}$~bar to high altitudes. The atmosphere structure modeling is of high computational cost (few hours), and therefore we restrict this analysis to a representative case rather than a full grid.

\subsection{Simulation Setup}

The initial planetary compositions used in this study follow the model of \citet{young_earth_2023}, with the addition of carbon in the metal phase. All C/O ratios reported in this study are given in mole units. The silicate phase of the core is composed of 94.7\% (molar) MgSiO$_3$, 3.3\%~MgO, 1.1\%~SiO$_2$, 0.7\%~Na$_2$O,  0.1\%~Na$_2$SiO$_3$, and minor amounts of FeO, and FeSiO$_3$. The initial envelope is hydrogen-dominated, consisting of 99.9\%~H$_2$ and 0.1\%~CO$_2$ by mole, corresponding to a solar C/O ratio of 0.5 (by moles) \citep{suarez-andres_co_2018}. The initial metal phase of the core is Fe-rich. In the models where we include C metal, we add 6.5\% (molar) C to the initial metal phase, corresponding to an initial core composition of 0.5~wt\% carbon and a C/O ratio of 0.02. This is different in the case when no reactive carbon in the metal is assumed which implies nearly zero wt\% carbon available for reaction in the cores and commensurately small core C/O ratio of near zero. In those models where we exclude C metal, the initial metal phase available for reaction is assumed to be 99.9\%~Fe by mole. In both setups, trace amounts of H and Si are also included in the metal phase. We choose these two scenarios as a means of isolating the role that carbon in metal has on the results, rather than the bulk carbon concentration of the core (i.e., there can be carbon sequestered in metal that is not reactive).

The total planetary mass ranges from 2 to 10~M$_\oplus$, and we explore atmospheric mass fractions from 0.1 to 10\%. For each configuration, we consider two thermal states: one with an AMOI temperature of 3000~K and a SME temperature of 3500~K, and another with an AMOI temperature of 4000~K and an SME temperature of 4500~K. The greater SME temperature is meant to simulate silicate-metal equilibration within the magma ocean but at modest depths, as expected for these under dense metals.

\section{Results} \label{sec:Results}

\begin{figure*}
    \centering
    \includegraphics[width=1\textwidth]{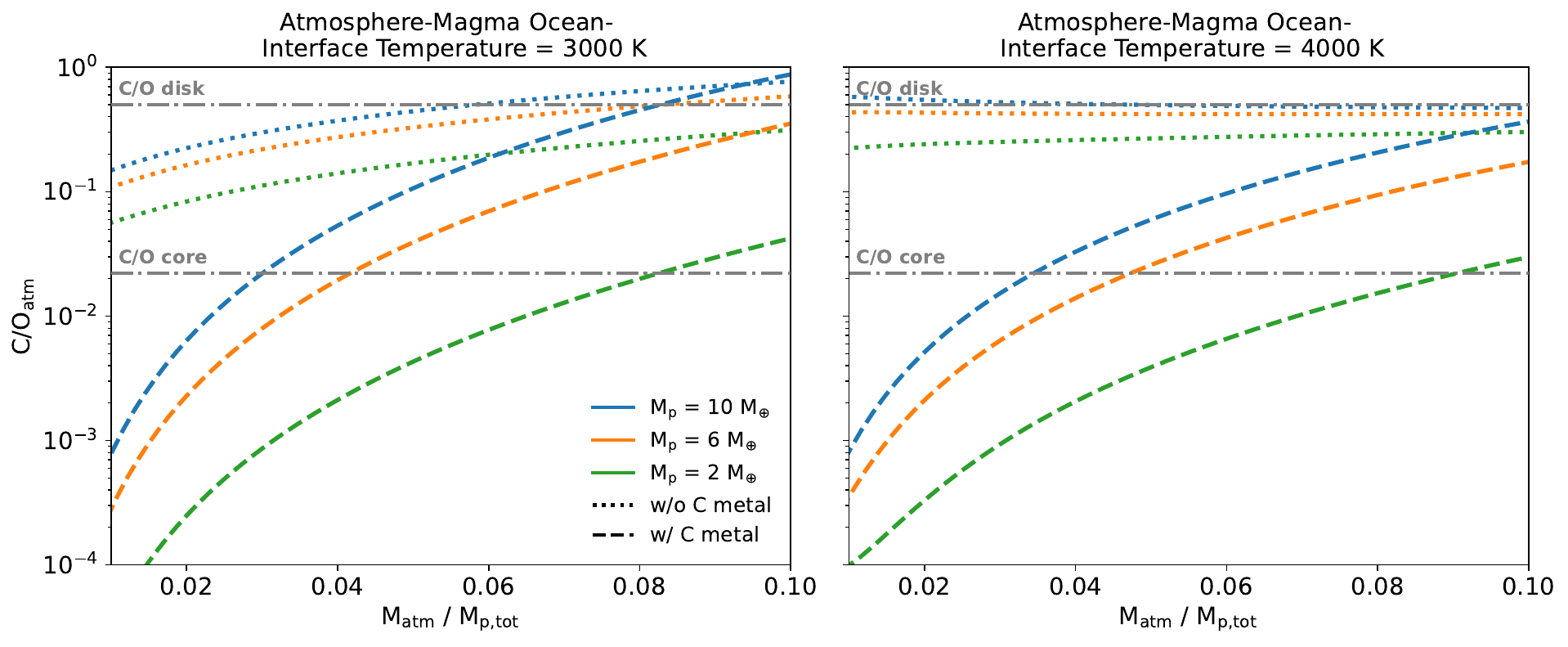}
    \caption{Molar bulk atmospheric C/O ratio as a function of atmospheric mass fraction. Dashed lines show results from the reaction network that includes carbon in the metal phase; dotted lines show results from the network excluding carbon in the metal phase. The horizontal dashed lines indicate the initial C/O ratios of the disk and the core. The core C/O ratio is calculated from the initial conditions for the case with C metal. The left panel shows models with a atmosphere–magma ocean interface (AMOI) temperature of 3000 K and a silicate–metal equilibrium (SME) temperature of 3500~K. The right panel shows models with a AMOI temperature of 4000~K and a SME temperature of 4500~K. Including carbon in the metal phase leads to a decrease in the atmospheric C/O ratio at low atmospheric mass fractions.}
    \label{fig:C_O_ratios}
\end{figure*}

\begin{figure*}
    \centering
    \includegraphics[width=1\textwidth]{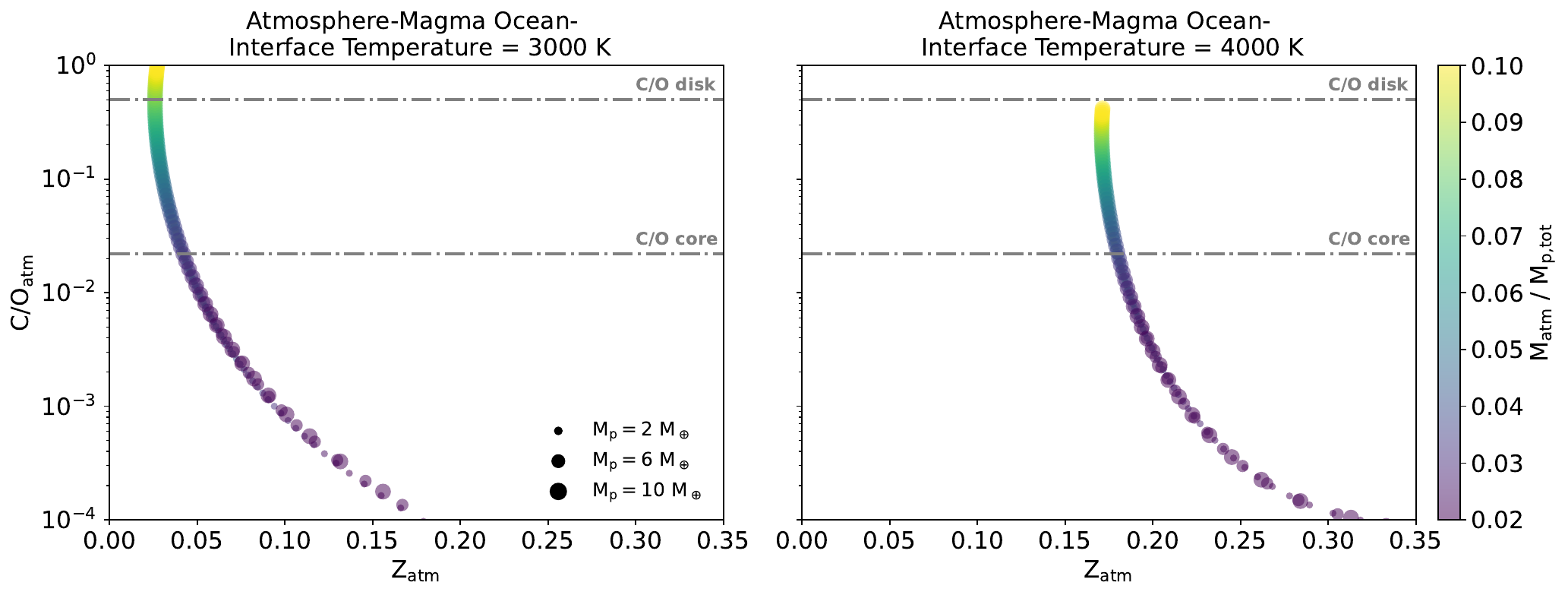}
    \caption{Molar bulk atmospheric C/O ratio as a function of resulting atmospheric metal mass fraction, $Z_{\text{atm}}$, for all models that include carbon in the metallic phase (from Figure~\ref{fig:C_O_ratios}). Horizontal dashed lines mark the initial C/O ratios of the disk and the core. The color bar indicates atmospheric mass fraction. Z$_{\text{atm}}$ is defined as the mass fraction of all atmospheric constituents except hydrogen, divided by the total atmospheric mass. Circle sizes denote different planetary masses. The left panel shows models with an atmosphere–magma ocean interface (AMOI) temperature of 3000 K and a silicate–metal equilibrium (SME) temperature of 3500~K; the right panel shows models with AMOI temperature of 4000~K and SME temperature of 4500~K. Across all cases, the atmospheric C/O ratio decreases with increasing Z$_{\text{atm}}$, independently of planetary mass.}
    \label{fig:metallicity}
\end{figure*}

\begin{figure}
    \centering
    \includegraphics{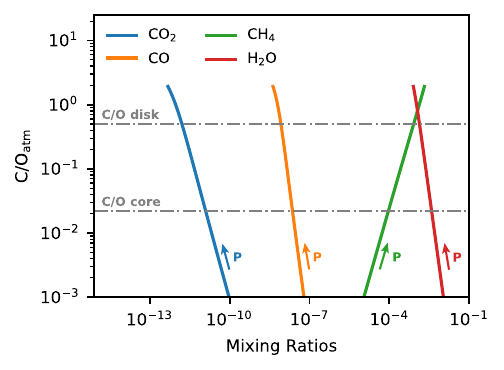}
    \caption{Molar bulk atmospheric C/O ratio as a function of species mixing ratios for the 3000K atmosphere–magma ocean interface (AMOI) temperature models with carbon in the metallic phase, from Figure~\ref{fig:C_O_ratios}. Horizontal dashed lines indicate the initial C/O ratios of the disk and the core. $P$ denotes AMOI pressure, which increases with increasing atmospheric C/O ratio. At high atmospheric mass fractions and planetary masses—corresponding to higher pressures—the atmosphere is carbon-rich and CH$_4$-dominated, driving up the C/O ratio. At low atmospheric mass fractions and low pressures, H$_2$O becomes more abundant, and the C/O ratio decreases.}
    \label{fig:mixing_ratios}
\end{figure}

\begin{figure}
    \centering
    \includegraphics{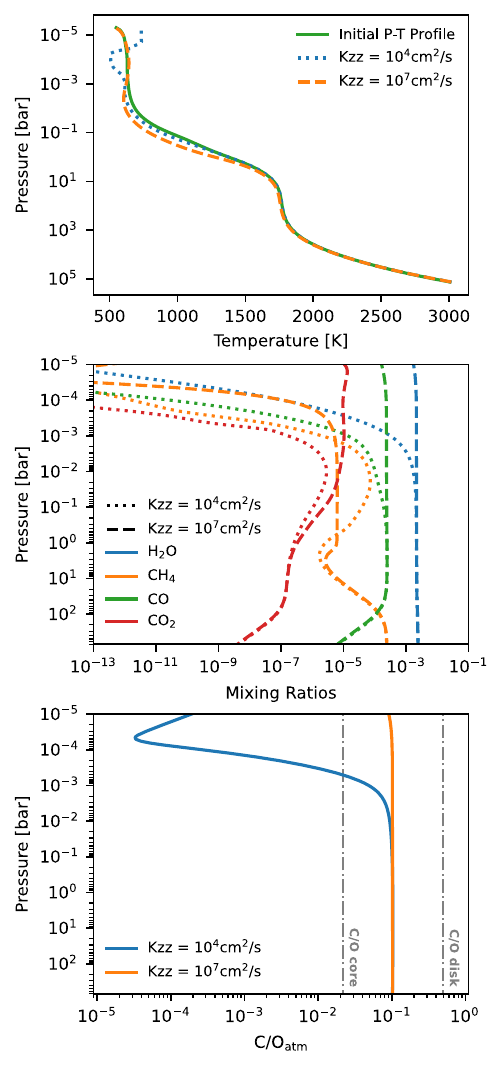}
    \caption{Atmospheric structure for a 6~M$_\oplus$ and 2.5~R$_\oplus$ planet with an atmosphere-magma ocean interface (AMOI) temperature of 3000~K and an equilibrium temperature of $\approx 600$~K. The radius is defined at 1 bar atmospheric pressure. All panels show results for two different eddy diffusion coefficients, K$_\mathrm{zz} = 10^4$ and $10^7$~cm$^2$s$^{-1}$, which control vertical mixing in the atmosphere. The stellar spectrum is modeled as a blackbody. The top panel shows the pressure–temperature profile before and after convergence; the middle panel shows the vertical variations in mixing ratios of the dominant carbon- and oxygen-bearing species; and the bottom panel shows the molar atmospheric C/O ratio as a function of pressure. Dashed vertical lines in the bottom panel indicate the initial C/O ratios of the disk and the core, highlighting the distinct C/O ratios of the atmosphere compare to core and disk. Lower vertical mixing rates lead to stronger vertical gradients in the atmospheric C/O ratio.}
    \label{fig:atmosphere_profiles}
\end{figure}

\subsection{C/O Ratio as a Function of Atmospheric Mass Fraction and Planetary Mass}

Figure~\ref{fig:C_O_ratios} shows the bulk atmospheric C/O ratio as a function of atmospheric mass fraction for planets with 2, 6, and 10~M$_\oplus$ and AMOI temperatures of 3000~K and 4000~K. The corresponding SME temperatures are 3500~K and 4500~K, respectively. Colored dashed lines represent simulations that include carbon in the metal phase, allowing carbon to partition into the metal via reaction \ref{Carbonreaction}. Colored dotted lines show results from simulations where carbon is excluded from the metal phase. The core (including silicates and metals) and disk C/O ratios are shown as horizontal grey dashed lines. The core C/O ratio is calculated from the initial conditions used in simulations, with 0.5 wt\% carbon in the core for the case with metal C, and zero in the C metal-free case. Therefore, the initial C/O for the core shown in Figure~\ref{fig:C_O_ratios} applies only to the cases with metal carbon included. 

In models that include carbon in the metal phase, the atmospheric C/O ratio decreases sharply at low atmospheric mass fractions, deviating by several orders of magnitude from the disk value. This effect is strongest for low-mass planets with small atmosphere mass fractions. At high atmospheric mass fractions, the C/O ratio approaches the initial disk value, as less carbon is sequestered into the core relative to the available volatile reservoir.

In contrast, models that exclude carbon in the metal phase yield significantly higher atmospheric C/O ratios, which remain nearly constant across all atmospheric mass fractions, particularly at the higher AMOI temperature of 4000~K. This behavior occurs because a higher SME temperature increases oxygen partitioning into the metal phase, leading to a depletion of oxygen in the atmosphere. Since a sink for carbon is absent, the amount of carbon in the atmosphere remains nearly constant. This combination results in an elevated and nearly constant C/O ratio.

The exact atmospheric C/O ratio depends on both the SME temperature and the adopted carbon partitioning fit. In particular, we find that increasing the SME temperature from 3500~K to 4000~K (for an AMOI temperature of 3000 K) can increase the atmospheric C/O ratio by up to two orders of magnitude for a given envelope mass fraction. However, the results are less sensitive to the choice of carbon partitioning fit: using the 1$\sigma$ uncertainty fit from \cite{blanchard_metalsilicate_2022} leads to only minor deviations in the resulting C/O ratios. While the absolute values depend on these parameters, the overall trends with planetary mass and atmospheric mass fraction remain robust. Figures illustrating the sensitivity analysis are provided in Appendix~\ref{ap:1}.

\subsection{C/O Ratio as a Function of Atmospheric Metal Mass Fraction}

Figure~\ref{fig:metallicity} shows the bulk atmospheric C/O ratio as a function of the atmospheric metal mass fraction, $Z_\mathrm{atm}$, where all atoms other than hydrogen are considered metals (our models do not include He). Horizontal dashed lines indicate the initial disk and core C/O ratios. Only models that include carbon in the metal phase are shown. Each data point corresponds to an individual planet from Figure~\ref{fig:C_O_ratios}. The left and right panels represent AMOI temperatures of 3000 K and 4000 K, respectively. Circle size denotes planetary mass, and the color bar indicates atmospheric mass fraction. Across all planetary masses and atmosphere mass fractions, the data collapse onto temperature-dependent sequences. At higher AMOI temperatures, enhanced evaporation of species like Mg, Si, Fe, and especially SiO leads to higher atmospheric metallicities and atmospheric mass fractions at a given C/O ratio. The increased abundance of SiO lowers the atmospheric C/O ratio, thereby shifting the entire distribution toward higher $Z_\mathrm{atm}$ and lower C/O.

In addition to that, the atmospheric C/O ratio decreases with increasing $Z_\mathrm{atm}$. This trend can best be understood by examining Figure~\ref{fig:mixing_ratios}, which shows the bulk atmospheric mixing ratios of the dominant carbon- and oxygen-bearing species as a function of atmospheric C/O ratio for the 3000~K case with carbon in the metallic phase. Horizontal dashed lines indicate the initial C/O ratios of the disk and the core. As the bulk atmospheric C/O ratio increases—corresponding to planets with higher masses, higher atmospheric mass fractions, and thus higher AMOI pressures—CH$_4$ becomes more abundant while H$_2$O decreases. These changes in molecular abundances explain the trend in atmospheric metal mass fraction seen in Figure~\ref{fig:metallicity}: at low $Z_\mathrm{atm}$ (high atmospheric mass fractions), the atmosphere is carbon-rich and CH$_4$-dominated, driving up the C/O ratio. At higher $Z_\mathrm{atm}$ (low atmospheric mass fractions), H$_2$O becomes dominant, lowering the C/O ratio. Overall, the atmospheric C/O ratio increases with atmospheric mass fraction and planetary mass (see Figure~\ref{fig:C_O_ratios}), and thus with AMOI pressure P.

\subsection{Vertical Transport and the Observability of the C/O Ratio}

Figure~\ref{fig:atmosphere_profiles} shows the atmospheric structure of a 6~M$_\oplus$ planet selected from the 3000 K AMOI temperature case shown in Figure~\ref{fig:C_O_ratios}. We assume a radius of 2.5~R$_\oplus$, defined at 1 bar atmospheric pressure. The equilibrium temperature is approximately 600~K, corresponding to an orbital distance of 0.1~AU around a 5200~K star. The stellar spectrum is modeled as a blackbody. We consider two vertical mixing scenarios with eddy diffusion coefficients of K$_\mathrm{zz} = 10^4$ and $10^7$~cm$^2$s$^{-1}$. The top panel shows the pressure–temperature (P–T) profile before and after convergence. The middle panel presents vertical variation in mixing ratios of the dominant oxygen- and carbon-bearing species from $10^3$~bar to $10^{-5}$~bar. In the low-mixing case, these mixing ratios drop significantly above $\sim$10$^{-2}$~bar, while they remain approximately constant (except for CH$_4$) in the high-mixing case. The bottom panel displays the atmospheric C/O ratio calculated from all carbon and oxygen bearing species. Vertical dashed grey lines indicate the initial C/O ratios of the disk and the core. In the high-mixing scenario, the C/O ratio remains constant with altitude, whereas in the low-mixing case it decreases above $\sim$10$^{-2}$ bar.

We calculate the molar C/O ratio using all carbon- and oxygen-bearing species output by VULCAN. In the low-mixing case, the observed drop in C/O ratio above $\sim 10^{-2}$ bar is primarily driven by gravitational settling. At these low pressures, heavier species with smaller scale heights—such as CO$_2$ (44 u) and CO (28 u)—settle more efficiently than lighter species like CH$_4$ (16 u) and H$_2$O (18 u) \citep{tsai_comparative_2021}. In our simulations, H$_2$O remains the dominant oxygen-bearing species throughout the envelope. As CH$_4$ overtakes CO as the main carbon carrier, the C/O ratio rises slightly at high altitudes.

Photochemistry also contributes to the depletion of CO, CH$_4$, and H$_2$O at low pressures, but its impact on the C/O ratio is secondary, as the abundances of photochemical products remain minor in our simulations. In general, the downward transport and eventual destruction of photochemical products may aid in removing some carbon and oxygen from the upper atmosphere \citep{tsai_comparative_2021}, but the effect on the C/O ratio is negligible.

In contrast, in the high-mixing case, efficient vertical transport continuously mixes carbon- and oxygen-bearing species throughout the atmosphere, preventing their depletion and maintaining a constant C/O ratio with altitude.

Typical observable regions in planetary atmospheres span pressure ranges of $10^{-1}$ to $10^{-3}$~bar for emission spectroscopy \citep{piette_rocky_2023} and $10^{-3}$ to $10^{-5}$~bar for transmission spectroscopy \citep{benneke_jwst_2024}. Both vertical mixing scenarios yield the same observable C/O ratio in emission, but in transmission spectroscopy the C/O ratio is affected by vertical mixing: in the low-mixing case, the C/O ratio decreases across observable pressures, while it remains constant in the high-mixing scenario.

\section{Discussion} \label{sec:Discussion}

We find that the atmospheric C/O ratio in sub-Neptunes is not inherited from the protoplanetary disk, but instead emerges from chemical equilibrium between the atmosphere and the underlying magma ocean. In itself, this result should  not be surprising, since most of the mass ($>$90\%) of sub-Neptunes investigated here resides in their magma-ocean and metal interiors. These interiors can therefore significantly modify the C/O ratio of the primordial hydrogen envelope that typically only contains a couple of percent of the total planet's mass. Specifically, we find that the C/O ratio increases with planetary mass and atmospheric mass fraction, yet remains distinct from both disk and core values. Including carbon in the metal phase reduces the atmospheric C/O ratio significantly, particularly for planets with low masses and low atmosphere mass fractions. The finding is particularly exciting, since it implies that observed C/O ratios of sub-Neptunes can be used to probe their interior composition and chemistry. Specifically, the detection of C/O ratios that are much lower than host star values, would imply an underlying magma-ocean with iron metal having sequestered a lot of the carbon.

Recent studies suggest that SiH$_4$ may be an important tracer of magma ocean activity \citep{misener_atmospheres_2023, charnoz_effect_2023} and could be detectable even at low pressures \citep{ito_monosilane_2025}. Although SiH$_4$ was not included in our main simulations, we conducted preliminary tests by adding it to the chemical network to explore its potential impact. These tests indicate that SiH$_4$ reduces the atmosphere and significantly promotes CH$_4$ formation, leading to elevated C/O ratios. However, since CH$_4$ is rapidly depleted within the envelope, high near-surface CH$_4$ abundances may not translate into elevated C/O ratios in observable regions. Exploring the full implications of SiH$_4$ is beyond the scope of this study but represents an important direction for future work.

Questions regarding magma ocean–atmosphere interactions have been explored by \citet{tian_atmospheric_2024} and \citet{seo_role_2024}, who model their influence on atmospheric composition without accounting for metal–silicate partitioning. \citet{tian_atmospheric_2024} explore a C–H–O system using imposed redox parameters and initial atmospheric compositions, and find that the atmospheric C/O ratio decreases with pressure. Their models predict CO and CO$_2$ as the dominant carbon-bearing species across a wide range of conditions, with CH$_4$ never emerging as the major component. In contrast, our chemical equilibrium model consistently yields CH$_4$ as the dominant carbon carrier in the deep atmosphere, and an increasing C/O ratio with pressure. While \citet{tian_atmospheric_2024} do not include photochemistry, we explicitly couple our interior compositions to 1D atmospheric models that incorporate vertical mixing and photochemical kinetics. For the atmosphere structure we analyzed, CH$_4$ is efficiently destroyed above $\sim$10$^{-2}$~bar (low K$_\mathrm{zz}$) and $\sim$10$^{-3}$~bar (high K$_\mathrm{zz}$), leading to a transition where CO and CO$_2$ become the dominant carbon-bearing species in the observable atmosphere. This transition is in agreement with the results of \citet{tian_atmospheric_2024}, though in our case it arises from photochemical depletion of CH$_4$ rather than equilibrium chemistry alone. In addition, \citet{tian_atmospheric_2024} also analyzes an extended C–H–O–N–S network, which lies outside the scope of direct comparison due to the absence of sulfur and nitrogen in our model. 

\citet{seo_role_2024} similarly study magma–gas equilibrium under various planetary conditions and find that O/H decreases with increasing C/O, a trend we also reproduce. While \citet{tian_atmospheric_2024} investigate AMOI temperatures between 1600–1800 K, \citet{seo_role_2024} explore atmospheric compositions up to 4000 K and derive the redox state of the mantle self-consistently from the assumed bulk composition.

Our work differs from both studies in two key aspects. First, we include a metal phase and allow carbon to partition into it, introducing a major sink that strongly modulates atmospheric composition. Second, we include a reaction between the metal and silicate phases that produces H$_2$O, providing an internal oxygen source that significantly influences the atmospheric C/O ratio.

Finally, we note that CH$_4$ is not included as a silicate species in our simulations, so direct dissolution into the melt is not explicitly accounted for. Given the low solubilities of CH$_4$ in silicate melts \citep{ardia_solubility_2013}, inclusion of an explicit reaction for CH4 dissolution is unlikely to have an impact on our results.

\section{Conclusion}\label{sec:Conclusions}

Our results demonstrate that atmospheric C/O ratios in sub-Neptunes are not inherited from the protoplanetary disk if there is a magma ocean beneath the dense H$_2$-rich atmosphere.  Under these circumstances, the atmospheric C/O ratio cannot be used directly as a tracer of planet formation location or disk chemistry, as is often assumed for hot Jupiters and some small icy Solar System bodies. Instead, the C/O ratio is significantly modified by chemical equilibrium between the atmosphere and the underlying magma ocean. We show that partitioning of carbon to the metallic phase further reduces the atmospheric C/O ratio, especially for low-mass planets with low atmosphere mass fractions. This internal processing decouples the atmospheric composition from its primordial origin. Importantly, we find that the deep C/O ratio is largely preserved in the observable atmosphere, opening up the opportunity to probe the interior composition and chemistry using spectroscopic observations of sub-Neptunes. For example, the detection of C/O ratios that are much lower than host star values, would imply an underlying magma-ocean with iron metal having sequestered a lot of the carbon.

These results suggest that atmospheric retrievals assuming disk-based C/O priors may lead to biased results. 

Our findings provide a framework in which to interpret the observed compositional diversity in sub-Neptunes and motivate future work exploring additional chemical species, time-dependent evolution, and application to JWST targets.


\section*{Acknowledgements}

C.D acknowledges support from the Swiss National Science Foundation under grant TMSGI2\_211313.
H.E.S gratefully acknowledges support from NASA under grant number 80NSSC18K0828. E.D.Y. acknowledges support from NASA grant number 80NSSC21K0477 issues through the Emerging Worlds program. This work has been carried out within the framework of the NCCR PlanetS supported by the Swiss National Science Foundation under grant 51NF40\_205606. We thank the anonymous reviewer for their insightful comments, which greatly helped to improve this study. We acknowledge the use of large language models (LLMs), including ChatGPT, to improve the grammar, clarity, and readability of the manuscript.

\section*{ORCID iDs}

\noindent Aaron Werlen \orcidlink{0009-0005-1133-7586} \href{https://orcid.org/0009-0005-1133-7586}{0009-0005-1133-7586} \\
Caroline Dorn \orcidlink{0000-0001-6110-4610} \href{https://orcid.org/0000-0001-6110-4610}{0000-0001-6110-4610} \\
Hilke E. Schlichting \orcidlink{0000-0002-0298-8089} \href{https://orcid.org/0000-0002-0298-8089}{0000-0002-0298-8089} \\
Simon L. Grimm \orcidlink{0000-0002-0632-4407} \href{https://orcid.org/0000-0002-0632-4407}{0000-0002-0632-4407} \\
Edward D. Young \orcidlink{0000-0002-1299-0801} \href{https://orcid.org/0000-0002-1299-0801}{0000-0002-1299-0801}

\appendix
\twocolumngrid

\section{Model Sensitivity Tests}\label{ap:1}

\begin{figure}
    \centering
    \includegraphics{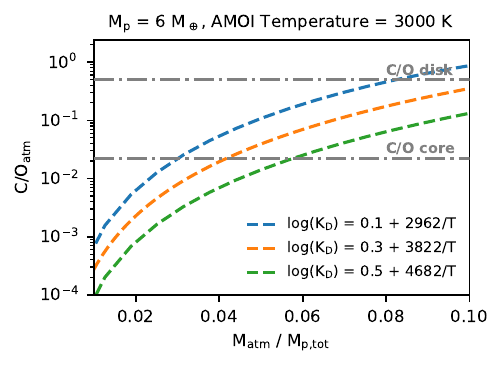}
    \caption{Molar bulk atmospheric C/O ratio as a function of atmospheric mass fraction. A 6~M$_\oplus$ planet with an AMOI temperature of 3000~K and an SME temperature of 3500~K is shown. The orange dashed line reproduces the default case from Figure~\ref{fig:C_O_ratios}. Two additional curves illustrate the effect of different carbon partitioning fits from \citet{blanchard_metalsilicate_2022}, spanning the 1$\sigma$ uncertainty range. Horizontal dashed lines indicate the initial C/O ratios of the disk and the core, with the core value derived from the initial conditions in the C-metal case.}
    \label{fig:sens_part}
\end{figure}

\begin{figure}
    \centering
    \includegraphics{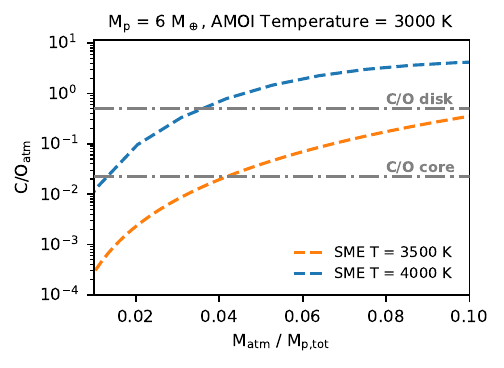}
    \caption{Same as Figure~\ref{fig:sens_part}, but showing the effect of varying SME temperature. Both curves use the same carbon partitioning fit as in Figure~\ref{fig:C_O_ratios}. The orange curve corresponds to the default case from Figure~\ref{fig:C_O_ratios}~\&~\ref{fig:sens_part}, while the blue curve corresponds to an SME temperature of 4000~K. Increasing the SME temperature leads to significantly higher C/O ratios due to enhanced oxygen partitioning into the metal phase.}
    \label{fig:sens_SME}
\end{figure}

To assess the sensitivity of our model to key assumptions, we re-ran the atmospheric C/O ratio calculations for a 6~M$_\oplus$ planet with varying atmospheric mass fraction, as in Figure~\ref{fig:C_O_ratios}. In Figure~\ref{fig:sens_part}, we explore the impact of different carbon partitioning fits, using alternative fits from \citet{blanchard_metalsilicate_2022} that span the 1$\sigma$ uncertainty range. In Figure~\ref{fig:sens_SME}, we test the effect of increasing the SME temperature from 500~K to 1000~K above the AMOI temperature.

We find that the partitioning fit only moderately shift the absolute C/O ratio. In contrast, the SME temperature has a significant effect, changing the atmospheric C/O ratio by up to two orders of magnitude. This strong sensitivity arises because higher SME temperatures increase the oxygen partitioning into the metal phase, reducing the oxygen available in the atmosphere. Despite these variations in absolute values, the qualitative trends and overall conclusions remain robust.

\FloatBarrier

\bibliography{references}{}
\bibliographystyle{aasjournal}



\end{document}